\documentclass[journal]{IEEEtran}
\usepackage{graphicx}
\usepackage{booktabs}
\usepackage{array}
\usepackage[normalem]{ulem}
\usepackage{amsmath,amssymb,bm,mathtools}
\usepackage{url}
\usepackage{xcolor}

\title{From Intraday Orderbook to Imbalance Price: Understanding \\ Cross-Market Interaction}
\author{Runyao Yu\textsuperscript{1,2,3},
Jochen L. Cremer\textsuperscript{2,3},
Pierre Pinson\textsuperscript{4},
Jalal Kazempour\textsuperscript{5}, \\
Leo Semmelmann\textsuperscript{6},
Takuji Matsumoto\textsuperscript{7},
and Derek W. Bunn\textsuperscript{1}%
\thanks{\textsuperscript{1}London Business School, \textsuperscript{2}Delft University of Technology, \textsuperscript{3}Austrian Institute of Technology, \textsuperscript{4}Imperial College London, \textsuperscript{5}Technical University of Denmark, \textsuperscript{6}Karlsruhe Institute of Technology, and
\textsuperscript{7}Institute of Science Tokyo.
}}

\begin{document}
\maketitle

\begin{abstract}
Power systems with increasing variable renewable generation face greater uncertainty in scheduling and balancing. Intraday and balancing electricity markets facilitate position adjustments and real-time balancing close to delivery. \textcolor{black}{As delivery approaches, continuous intraday market participants exposed to imbalance settlement adjust their positions by trading additional volumes to reduce their imbalance exposure. We conjecture that positions remaining open after intraday trading, together with demand and supply uncertainties affecting physical market participants, influence price formation in the balancing market.} This cross-market interaction is, however, rarely studied. 
\textcolor{black}{To understand this interaction, this paper uses probabilistic modeling to examine how intraday orderbook information reflects subsequent imbalance price formation in Germany and Austria.}
We compare orderbook representations based on open, high, low, close, and volume, Volume-Weighted Average Price (VWAP), and last mid price across multiple horizons. Each representation is evaluated using the self product, neighboring products, and the product from the neighboring country. We then compare the best orderbook setting with fundamental feature sets and their combinations, followed by an ablation study of the available training history. We show that VWAP with neighboring products provides the best performance in both countries. Combining orderbook and fundamental information reduces testing loss in Germany but increases it in Austria. Using all available observations provides the best overall performance, while excluding 2022 can reduce loss for extreme price samples. These results reveal and help explain country-dependent interactions between the intraday and balancing markets.
\end{abstract}

\section{Introduction}
Power systems are transitioning toward variable renewable generation, whose variability, stochasticity, and limited controllability increase balancing challenges \cite{pinson2023distributionally, WANG2026104781, CHEN2026128554}. Electricity markets coordinate resources to address these challenges \cite{11569340}. In sequence, the three main short-term electricity markets are the day-ahead market, the intraday market, and the balancing market \cite{11589893}. Most volume is traded in the day-ahead market, after which renewable generation and load forecast errors create scheduling deviations. Subsequently, intraday trading adjusts positions to respond to the renewable forecast errors. The balancing market settles the resulting aggregate net deviations from the system operator's target at the grid level. All participants are then exposed to their individual real-time deviations from nominations at the imbalance settlement price. The resulting imbalance price is particularly volatile close to delivery, as it depends on the system operator's short-term imbalance forecast and the marginal prices of the various reserve products activated by the system operator \cite{11150599}.


\begin{figure*}[t]
\centering
\includegraphics[width=1.01\textwidth]{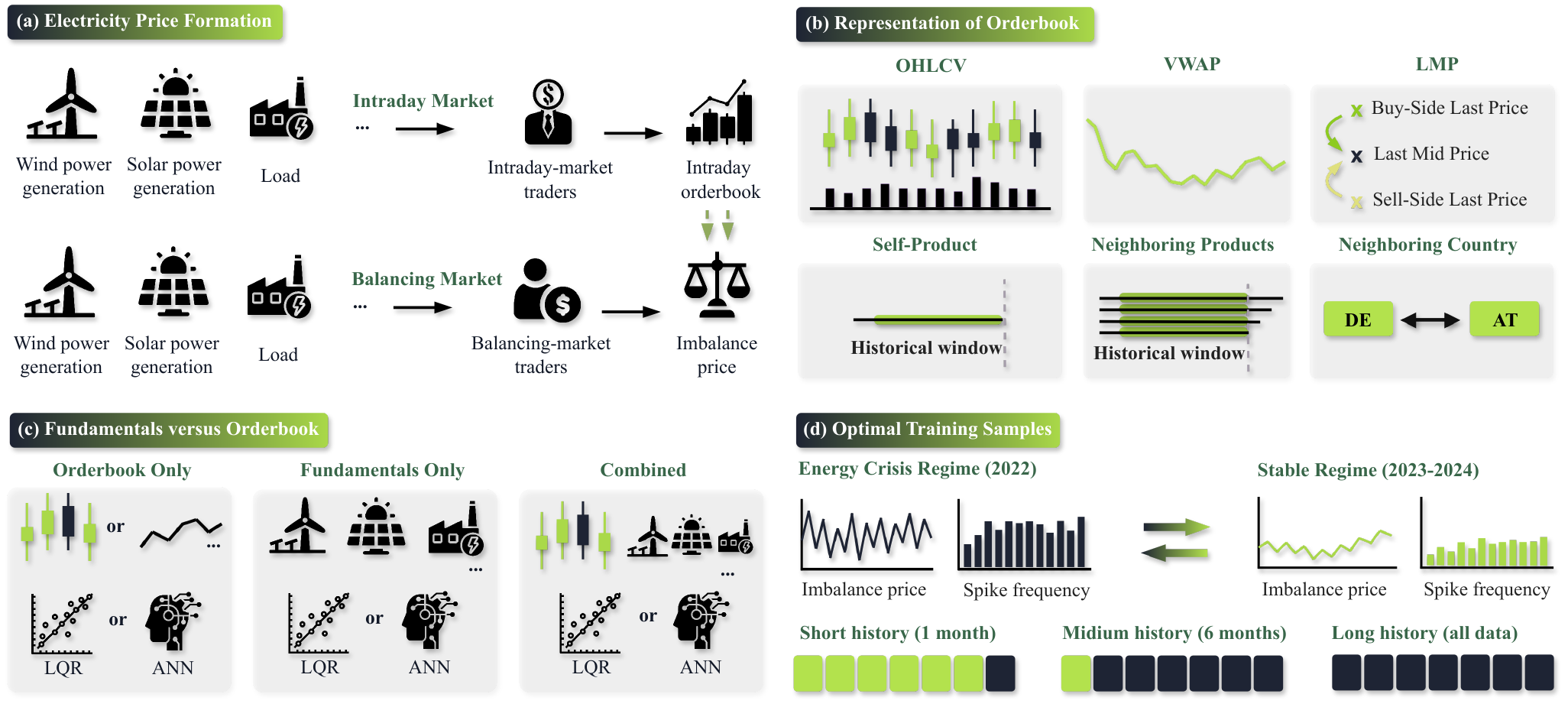}
\caption{Overview of the work. \textbf{(a)} From fundamentals to electricity prices. Continuous intraday traders adjust their orders in response to renewable generation and load, forming the intraday orderbook, while the imbalance price also reflects fundamental supply--demand in balancing markets. \textbf{(b)} Three representations of the orderbook (OHLCV, VWAP, and LMP) and their variants. \textbf{(c)} Three feature case studies (orderbook only, fundamentals only, and their combination) evaluated using LQR and ANN models. \textbf{(d)} Investigation of the optimal training sample size by varying the amount of historical training data.}
\label{fig:overview}
\end{figure*}

Accurate modeling of future imbalance prices can support power system management \cite{ganesh2023forecasting} and increase trading revenue \cite{natarajan2024algorithmic}. In markets such as Belgium and the United Kingdom, market participants with flexible assets, for example Battery Energy Storage Systems (BESS), can participate across \textcolor{black}{several} market stages \cite{semmelmann2026quantile}, including speculation upon their imbalance settlement exposures. Speculation on imbalance exposure is usually profitable, where it is allowed, if there is a single imbalance price and the participant can take an open position in the opposite direction to the balancing market. Thus, if the participant expects the balancing market to be net short, i.e., the system operator buying, it can deliberately go long by buying in the intraday market to be subsequently compensated for that imbalance. Similarly, if the participant expects the balancing market to be net long, it can deliberately go short. Therefore, providing the participant's optimizer with information about the future imbalance price can improve its market decisions and revenue \cite{miskiw2025continuous}, \mbox{\cite{SCHAURECKER2026128550}}.

Price modeling is well established in day-ahead and intraday electricity markets. However, imbalance price forecasting has received less attention \cite{11150599}, and whilst our focus is upon the information flow between markets and the price formation process, insights from previous work on forecasting help to inform the relevant drivers. Day-ahead studies apply graph neural networks \cite{yang2024forecasting} and foundation models \cite{yu2026pricefmfoundationmodelprobabilistic} to forecast the next day's electricity price and support scheduling. In the intraday electricity market, most studies apply linear regression \cite{Trading,Simulation} to forecast short-term prices and help reduce imbalance pressure. Other studies apply tree models \cite{YU2027113596}, and one study designs a novel order fusion model incorporating market microstructure \cite{yu2026orderfusionencodingorderbookendtoend}. For the balancing market, transformer variants \cite{ganesh2023forecasting} and market-rule-informed neural networks \cite{YU2026105083} improve computational efficiency. These imbalance price forecasting models rely on fundamentals such as renewable generation, load, and power flows.

Fig.~\ref{fig:overview} \textbf{(a)} illustrates how fundamentals such as renewable generation and load influence the intraday orderbook and imbalance price as established market context. This forms an implicit interaction between the intraday and balancing markets~\cite{koch2019short}. On the one hand, intraday prices become more volatile near delivery as traders reflect on fundamentals and seek to close their positions. 
This \textcolor{black}{is} because entering the balancing market might introduce imbalance price penalties \cite{bobo2018offering}. On the other hand, residual power positions left unclosed in the intraday market contribute to extreme imbalance prices. 
\textcolor{black}{Thus, this raises a key question: to what extent does the intraday orderbook reflect the information underlying subsequent imbalance price formation?}
This question matters in practice. The orderbook is a single signal that is observable in real time, whereas fundamental variables come from many sources and are partly published with delay. If the orderbook already aggregates the relevant information, market participants and system operators can anticipate imbalance prices from the orderbook alone. If not, the gap shows which physical information the intraday market fails to price in.

\begin{figure*}[!t]
\centering
\includegraphics[width=\textwidth]{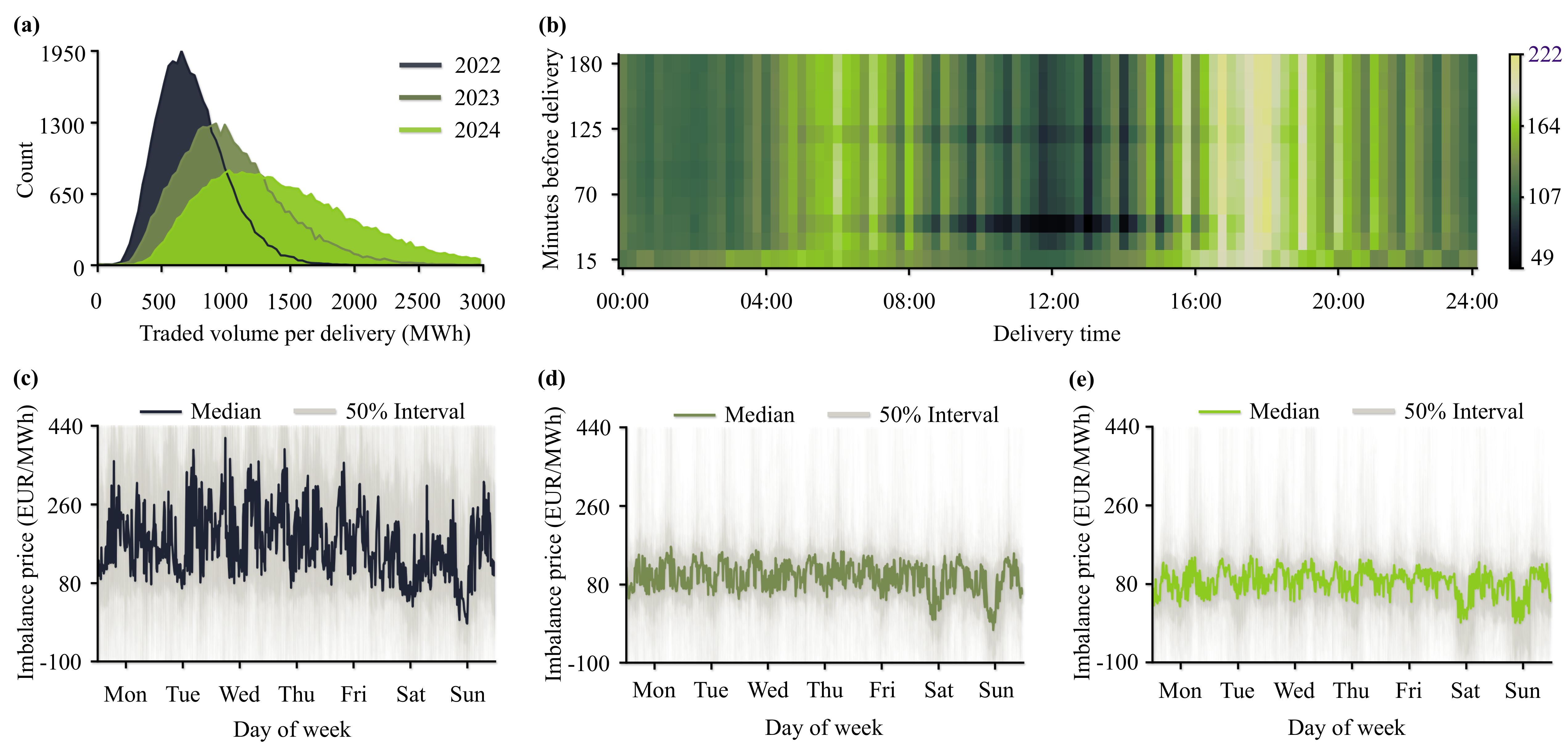}
\caption{Exploratory data analysis for Germany. Panel \textbf{(a)} shows annual traded volume distributions by product in the continuous intraday market. Panel \textbf{(b)} illustrates continuous intraday market VWAP by delivery time and minutes before delivery. Panels \textbf{(c)} to \textbf{(e)} overlay complete imbalance price weeks for 2022, 2023, and 2024. The line is the pointwise median and the band is the middle 50\% of weeks.}
\label{fig:deeda}
\end{figure*}

\begin{figure*}[!t]
\centering
\includegraphics[width=\textwidth]{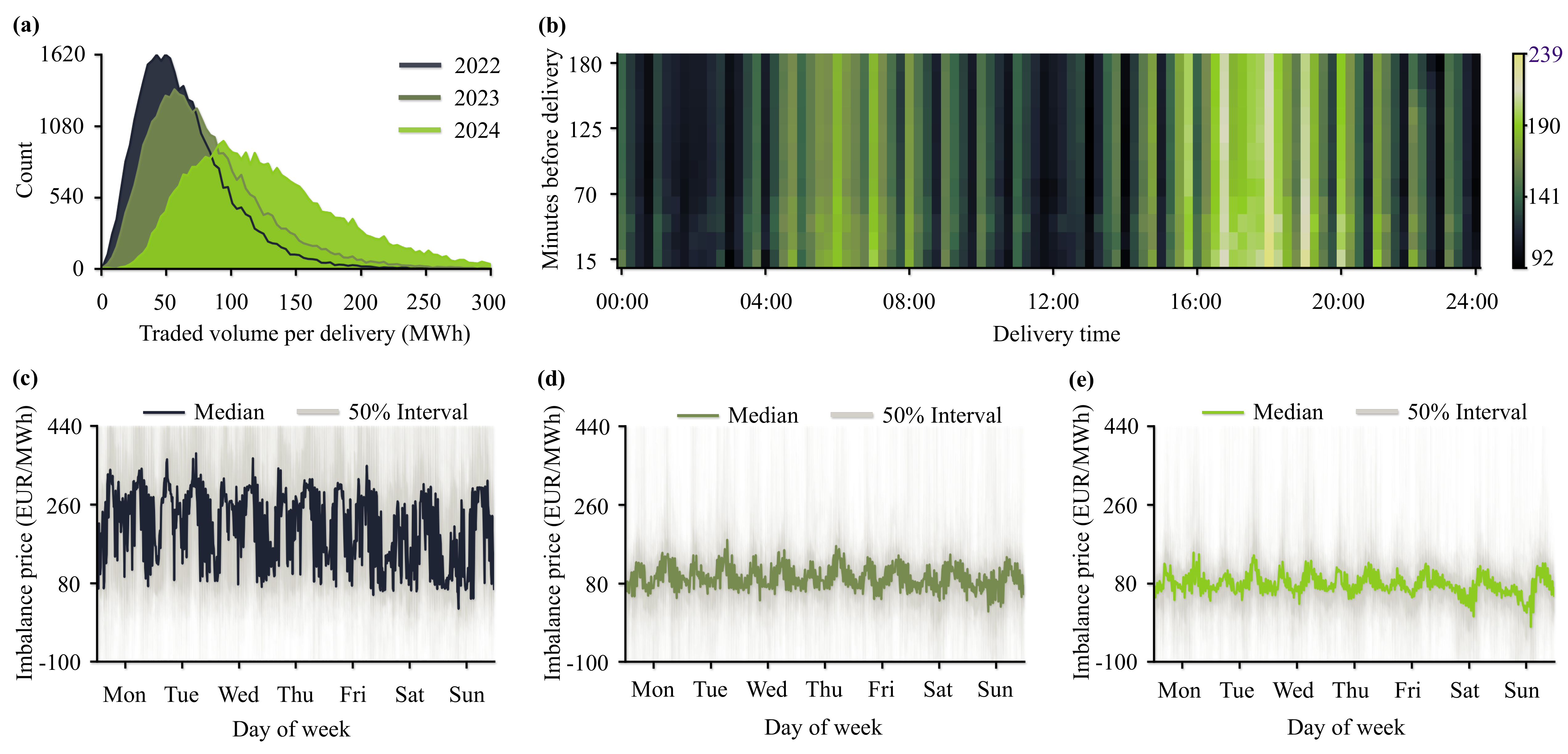}
\caption{Exploratory data analysis for Austria. Panel order, aggregation, and yearly definitions follow Fig.~\ref{fig:deeda}.}
\label{fig:ateda}
\end{figure*}

As raw orderbook observations are irregularly sampled, prior studies extract technical features to represent orderbook data. In particular, high and low prices \cite{YU2027113596} and Volume-Weighted Average Price (VWAP) \cite{beating, Trading, yu2026orderfusionprobabilisticbuysellprice} are important features, while other works indicate weak-form efficiency and suggest that the latest transaction price may suffice \cite{Simulation, Understanding}. However, the most suitable orderbook representation remains unknown for imbalance price modeling.

Existing studies also rarely examine the amount of historical data used for model training and commonly use all available past observations. Electricity price distributions can shift over time \cite{lima2022bayesian, johnsen2025electricity}, with the exceptional prices during the 2022 European energy crisis providing a prominent example. It is therefore unclear whether including the full historical sample improves imbalance price modeling.

In this paper, we focus on Germany and Austria, as the two countries offer a natural contrast. They formed a common bidding zone until 2018 and remain coupled through the European continuous intraday market, yet they differ strongly in scale and liquidity. Germany is a large and liquid market with a high share of wind and solar generation, whereas Austria is a much smaller market with roughly one tenth of the German traded volume. The information content of the orderbook and the added value of fundamentals may therefore differ between the two countries.

We compare representations based on Open, High, Low, Close, and Volume (OHLCV), VWAP, and Last Mid Price (LMP). Using the best orderbook representation, we compare orderbook models with fundamental models and examine whether the two information sources should be combined. An orderbook model uses only features extracted from the intraday orderbook and thus reflects the aggregated expectations that traders reveal through prices and volumes. A fundamental model uses only physical system variables, such as renewable generation, load, and cross-border exchange, which are the conventional inputs in the imbalance price literature. We hypothesize that the two sources can be complementary. The orderbook contains information processed by traders, including private information, whereas fundamentals describe the physical system state that traders may not have fully priced in. If the orderbook already prices in the fundamentals, combining them adds only redundant inputs. We then vary the training length and the inclusion of 2022 observations, as illustrated in Fig.~\ref{fig:overview}. Our contributions are:
\begin{itemize}
\item To the best of our knowledge, this is the first study to explore the representation of continuous intraday orderbook information for probabilistic imbalance price modeling.
 \item We conduct systematic comparisons and show that Austrian orderbook information is sufficient without fundamental features, whereas German orderbook information benefits from combination with fundamentals. 
\item We design training sample ablation studies and reveal that adding the 2022 energy crisis observations improves overall accuracy, while, surprisingly, excluding these observations can reduce loss for extreme imbalance price samples.
\end{itemize}

The remainder of this paper is organized as follows. Section~II describes the data. Section~III presents the models, orderbook representations, feature sets, and evaluation metrics. Section~IV reports the three case studies. Section~V discusses the limitations and future work, and Section~VI concludes the paper.

\section{Data}
The study uses continuous intraday orderbook data from EPEX Spot and fundamental power system data (renewable generation forecasts and actuals, load forecasts and actuals, day-ahead prices, cross-border net positions, and system imbalance, detailed in Section~III-C) and imbalance prices from the European Network of Transmission System Operators for Electricity (ENTSO-E) for Germany and Austria from January 2022 through January 2025. 

Fig.~\ref{fig:deeda} and Fig.~\ref{fig:ateda} summarize the German and Austrian data. Panel \textbf{(a)} in both figures shows traded volume per product increasing and shifting higher from 2022 to 2024, with the German market approximately 10 times the Austrian market by volume. This difference in scale and liquidity is one reason why the two case studies may yield different results. Panel \textbf{(b)} shows that VWAP is generally higher near 06:00 and 18:00 and lower near 12:00. Germany also exhibits repeated lower price bands from 120 to 105 minutes and from 45 to 30 minutes before delivery, which may reflect algorithmic trading schedules. Panels \textbf{(c)} to \textbf{(e)} show that the weekly imbalance price profile becomes flatter and less dispersed from 2022 to 2024. The wider 2022 distribution is consistent with the exceptional price conditions during the European energy crisis.

\section{Method}
Let $\mathcal{C}=\{\mathrm{DE},\mathrm{AT}\}$ denote the two countries and let $\mathcal{H}=\{15,60,180\}$ minutes denote the modeling horizons from extreme short term to short term. The input length $M\in\mathcal{M}=\{15,60,180\}$ minutes is selected as a hyperparameter for each horizon and model. The quantile set is $\mathcal{Q}=\{0.1,0.5,0.9\}$. At modeling origin $t$, $\bm{F}^{c}_{t-M:t}$ denotes historical inputs for country $c$ that can contain any feature, such as VWAP, and $\bm{F}^{c}_{t:t+H|t}$ denotes available lead inputs for country $c$ such as day-ahead and intraday solar generation forecasts, i.e., forecasts for the period from $t$ to $t+H$ that are issued and available at $t$. Hence, only information available at the origin $t$ is used. The availability of each fundamental variable at the origin is discussed in Appendix~A. For country $c\in\mathcal{C}$, horizon $H\in\mathcal{H}$, and quantile $q\in\mathcal{Q}$, a model $f_{\bm{\theta}}(\cdot)$ with parameters $\bm{\theta}$ produces
\begin{equation}
\widehat{y}^{(q)}_{c,t+H}
=f_{\bm{\theta}}^{(q)}\!\left(\bm{F}^{c}_{t-M:t},\bm{F}^{c}_{t:t+H|t}\right),
\label{eq:model}
\end{equation}
where $\widehat{y}^{(q)}_{c,t+H}$ is the modeled $q$-quantile of the imbalance price $y_{c,t+H}$ in country $c$ for delivery time $t+H$.
\begin{table}[!t]
\caption{Rolling origin data splits.}
\label{tab:folds}
\centering
\scriptsize
\setlength{\tabcolsep}{4.4pt}
\begin{tabular*}{\columnwidth}{@{\extracolsep{\fill}}llll}
\toprule
Fold & Training & Validation & Testing \\
\midrule
1 & Jan. 2022 to Aug. 2023 & Sep. to Dec. 2023 & Jan. to Apr. 2024 \\
2 & Jan. 2022 to Dec. 2023 & Jan. to Apr. 2024 & May to Aug. 2024 \\
3 & Jan. 2022 to Apr. 2024 & May to Aug. 2024 & Sep. to Dec. 2024 \\
\bottomrule
\end{tabular*}
\end{table}
\subsection{Models}
The function $f_{\bm{\theta}}(\cdot)$ can be any quantile regression model. In this paper, we consider two, namely an Artificial Neural Network (ANN) for nonlinear modeling and Linear Quantile Regression (LQR) for linear modeling. Our aim is to compare information sets and not to find the best modeling architecture. We therefore choose a generic and widely used nonlinear learner together with its linear counterpart. With all inputs kept identical, this pair isolates the value of nonlinearity.

\textbf{ANN.} ANNs are widely used for electricity price modeling \cite{o2025optimising}. According to the universal approximation theorem, an ANN with one sufficiently wide hidden layer can approximate any continuous function on a compact domain \mbox{\cite{hornik1989multilayer}}, \cite{lu2021learning}. In our setting, this means that an ANN can capture nonlinear relations between orderbook features and the imbalance price without specifying them a priori. We consider the Swish activation as prior work reports improved performance over conventional activation functions \cite{ramachandran2017searching}. Swish is smooth and non-monotonic and keeps a small gradient for negative inputs, which avoids inactive neurons and eases optimization. The activation function, number of hidden layers, number of neurons per layer, and learning rate are selected as hyperparameters on the validation set of each fold (Table~\ref{tab:folds}), with the search ranges listed in Table~\ref{tab:hyper} in Appendix~B. Each model is trained for 100 epochs, and only the weights that achieve the lowest validation loss within these epochs are retained for testing.

\textbf{LQR.} The aim of using LQR is to provide a linear benchmark for the nonlinear ANN and reveal whether the input features have an approximately linear relationship with the imbalance price. If LQR matches ANN performance, the simpler LQR is preferable, as it is interpretable, cheaper to estimate, and less prone to overfitting \cite{domingos1999role}.

\subsection{Orderbook Representation}
The orderbook is resampled into 15-minute intervals for each delivery time $t_d$, which is a common choice for orderbook representation \cite{hirsch2024multivariate}.

 \textbf{OHLCV.} For side $s\in\{\mathrm{buy},\mathrm{sell}\}$, let $\mathcal{I}_{t_d,k}^{s}$ be the ordered index set for the price and volume observations $(p_i^s,v_i^s)$ in interval $k$. 
 Missing intervals, \textcolor{black}{which occur when no orderbook observation is recorded for a delivery product within a 15-minute interval,} are masked with zeros. They are never filled over time or replaced with future information.
 The side-specific OHLCV representation is
\begin{equation}
\begin{aligned}
O_{t_d,k}^{s}&=p_{\min \mathcal{I}_{t_d,k}^{s}}^{s}, &
H_{t_d,k}^{s}&=\max_{i\in\mathcal{I}_{t_d,k}^{s}}p_i^{s},\\
L_{t_d,k}^{s}&=\min_{i\in\mathcal{I}_{t_d,k}^{s}}p_i^{s}, &
C_{t_d,k}^{s}&=p_{\max \mathcal{I}_{t_d,k}^{s}}^{s},\\
V_{t_d,k}^{s}&=\sum_{i\in\mathcal{I}_{t_d,k}^{s}}v_i^{s}.&&
\end{aligned}
\label{eq:ohlcv}
\end{equation}
Here, $\min\mathcal{I}_{t_d,k}^{s}$ and $\max\mathcal{I}_{t_d,k}^{s}$ are the first and last indices of the time-ordered set, so the open and close are the earliest and latest prices observed in interval $k$, while the high and low are taken over all prices in the interval.

 \textbf{VWAP.} 
 The price and volume observations $(p_i^s,v_i^s)$ are executed transactions. Hence, $v_i^s$ is the traded volume of trade $i$, and each trade enters the representation once. \textcolor{black}{VWAP provides a robust summary of the prevailing orderbook price level by assigning greater influence to trades with larger volumes and reducing the effect of low-volume price fluctuations.}
 In contrast, OHLCV describes the range and end points of the prices and carries volume only as a separate total, so a single small order at an extreme price can set the high or low. OHLCV thus captures price dispersion within the interval, whereas VWAP captures the price level at which most volume is traded. The two are evaluated as alternative representations and are not used jointly.
 The side-specific VWAP representation is
\begin{equation}
\operatorname{VWAP}_{t_d,k}^{s}
=\frac{\sum_{i\in\mathcal{I}_{t_d,k}^{s}}p_i^{s}v_i^{s}}
{\sum_{i\in\mathcal{I}_{t_d,k}^{s}}v_i^{s}}.
\label{eq:vwap}
\end{equation}

 \textbf{LMP.} LMP is the mean of the latest buy and sell closes and is given by
\begin{equation}
\operatorname{LMP}_{t_d,k}
=\frac{C_{t_d,k}^{\mathrm{buy}}+C_{t_d,k}^{\mathrm{sell}}}{2}.
\label{eq:lmp}
\end{equation}

For each representation, the self-product setting (Self) uses only the historical orderbook for the product delivered at $t_d$, i.e., the product whose imbalance price is modeled. The neighboring product settings add the same historical window for the next 1, 4, or 12 products, which are denoted by $+1$, $+4$, and $+12$. For example, if Self uses the orderbook for the 18:00 product, then $+1$ also uses the orderbook for the 18:15 product. Accordingly, $+4$ uses the orderbooks for the four later products from 18:15 through 19:00.
The neighboring products are included as a prior study reports that incorporating information from neighboring products improves future intraday price modeling \cite{hirsch2024multivariate}. We investigate whether this finding also holds for imbalance price modeling. The neighboring-country setting, in contrast, includes the product with the same delivery time from the other country.

\subsection{Feature Sets}
Let $\bm{x}^{r,c}_{t_d,k}$ collect the buy and sell side features of representation $r\in\{\mathrm{OHLCV},\mathrm{VWAP},\mathrm{LMP}\}$ for the product delivered at $t_d=t+H$ in country $c$ and interval $k$, i.e., ten values for OHLCV in \eqref{eq:ohlcv}, two for VWAP in \eqref{eq:vwap}, and one for LMP in \eqref{eq:lmp}. Let $\mathcal{K}_{t,M}$ be the intervals within the input window from $t-M$ to $t$, let $\Delta=15$ minutes, and let $c'$ be the neighboring country. The orderbook feature sets are
\begin{equation}
\begin{aligned}
\mathcal{F}^{r\text{-Self}}_{c}&=\{\bm{x}^{r,c}_{t_d,k}: k\in\mathcal{K}_{t,M}\},\\
\mathcal{F}^{r\text{-Neighbor}}_{c}&=\{\bm{x}^{r,c}_{t_d+j\Delta,k}: k\in\mathcal{K}_{t,M},\, j=0,\dots,n\},\\
\mathcal{F}^{r\text{-Country}}_{c}&=\mathcal{F}^{r\text{-Self}}_{c}\cup\mathcal{F}^{r\text{-Self}}_{c'},
\end{aligned}
\label{eq:featuresets}
\end{equation}
where $n\in\{1,4,12\}$ is the number of later products, corresponding to the $+1$, $+4$, and $+12$ settings of Section~III-B, and is selected on validation data. The fundamental feature sets are nested, $\mathcal{F}^{\text{Fund-}1}_{c}\subset\mathcal{F}^{\text{Fund-}2}_{c}\subset\mathcal{F}^{\text{Fund-}3}_{c}\subset\mathcal{F}^{\text{Fund-}4}_{c}$, and are listed in Table~\ref{tab:fundamental_sets}. Each set adds a further layer of information about the physical system state. Fundamental-1 contains the renewable generation forecasts and actuals that drive most forecast errors, Fundamental-2 adds the demand side through load forecasts and actuals, Fundamental-3 adds the day-ahead price as the market clearing of the scheduled system, and Fundamental-4 adds the cross-border net positions and the operational system imbalance that describe the physical state closest to delivery. Actual values enter the historical inputs $\bm{F}^{c}_{t-M:t}$ and forecasts enter the lead inputs $\bm{F}^{c}_{t:t+H|t}$. An orderbook model uses one orderbook set, a fundamental model uses one $\mathcal{F}^{\text{Fund-}j}_{c}$, and a combined model uses their union. We write, for example, VWAP-Neighbor and Fundamental-$j$ for the corresponding models.

\begin{table}[!t]
\caption{Cumulative fundamental feature sets.}
\label{tab:fundamental_sets}
\centering
\scriptsize
\setlength{\tabcolsep}{2.4pt}
\begin{tabular*}{\columnwidth}{@{\extracolsep{\fill}}p{0.62\columnwidth}cccc}
\toprule
 & \multicolumn{4}{c}{ Fundamental-} \\
Feature & 1 & 2 & 3 & 4 \\
\midrule
Solar and wind day-ahead forecasts & $\checkmark$ & $\checkmark$ & $\checkmark$ & $\checkmark$ \\
Solar and wind intraday forecasts & $\checkmark$ & $\checkmark$ & $\checkmark$ & $\checkmark$ \\
Actual solar and wind generation & $\checkmark$ & $\checkmark$ & $\checkmark$ & $\checkmark$ \\
Day-ahead load forecast & & $\checkmark$ & $\checkmark$ & $\checkmark$ \\
Actual load & & $\checkmark$ & $\checkmark$ & $\checkmark$ \\
Day-ahead price & & & $\checkmark$ & $\checkmark$ \\
Actual cross-border net position & & & & $\checkmark$ \\
Scheduled cross-border net position & & & & $\checkmark$ \\
Operational system imbalance & & & & $\checkmark$ \\
\bottomrule
\end{tabular*}
\end{table}

\subsection{Evaluation Metrics}
\textcolor{black}{We evaluate probabilistic performance using Average Quantile Loss (AQL) and Average Quantile Coverage Error (AQCE), pointwise performance using Mean Absolute Error (MAE) and Root Mean Squared Error (RMSE), and statistical differences using paired Diebold-Mariano (DM) tests. } The four accuracy metrics are complementary. AQL scores the full set of quantile predictions and is the training objective \cite{yu2026pricefmfoundationmodelprobabilistic, yu2026orderfusionencodingorderbookendtoend}, AQCE isolates the calibration of the 80\% prediction interval \cite{yu2026orderfusionencodingorderbookendtoend}, MAE measures the typical error of the median prediction, and RMSE emphasizes large errors and thus the extreme imbalance prices that matter most for market participants \cite{hyndman2006another}. The DM test is used to check whether a difference in any of these metrics is statistically significant and not due to sampling noise.

\textbf{AQL.} Let $N$ denote the number of samples and let $\mathcal{Q}$ denote the set of quantiles. The quantile loss is $\rho_q(u)=u\left(q-\mathbb{I}\{u<0\}\right)$, where $\mathbb{I}$ is the indicator function. AQL is defined as
\begin{equation}
\operatorname{AQL}
=
\frac{1}{N|\mathcal{Q}|}
\sum_{i=1}^{N}
\sum_{q\in\mathcal{Q}}
\rho_q\left(y_i-\widehat{y}_i^{(q)}\right).
\label{eq:aql}
\end{equation}
A lower AQL indicates more accurate probabilistic predictions.

\section{Case Study}
\begin{table*}[!t]
\caption{Orderbook testing performance across the horizons in Germany.}
\label{tab:orderbook_horizon_de}
\centering
\setlength{\arrayrulewidth}{\lightrulewidth}
\renewcommand{\arraystretch}{1.04}
\makebox[\textwidth][c]{%
\begin{tabular}{lc|c|c|c}
\toprule
Method & \begin{tabular}{@{}rrrr@{}}\multicolumn{4}{c}{\uline{Horizon $h$ = 15 min}}\\AQL & AQCE & MAE & RMSE\end{tabular} & \begin{tabular}{@{}rrrr@{}}\multicolumn{4}{c}{\uline{ Horizon $h$ = 60 min}}\\AQL & AQCE & MAE & RMSE\end{tabular} & \begin{tabular}{@{}rrrr@{}}\multicolumn{4}{c}{\uline{ Horizon $h$ = 180 min}}\\AQL & AQCE & MAE & RMSE\end{tabular} & Count \\
\midrule
Orderbook OHLCV-Self & \begin{tabular}{@{}rrrr@{}}30.03 & \textbf{0.026} & 91.69 & 236.96\end{tabular} & \begin{tabular}{@{}rrrr@{}}31.12 & \textbf{0.022} & 94.19 & 264.61\end{tabular} & \begin{tabular}{@{}rrrr@{}}34.24 & \textbf{0.019} & 100.94 & 277.89\end{tabular} & 3/12 \\
Orderbook OHLCV-Neighbor & \begin{tabular}{@{}rrrr@{}}28.48 & \textbf{0.042} & 85.47 & 207.95\end{tabular} & \begin{tabular}{@{}rrrr@{}}32.28 & 0.052 & 96.30 & 261.21\end{tabular} & \begin{tabular}{@{}rrrr@{}}33.88 & \textbf{0.047} & 97.41 & 266.68\end{tabular} & 2/12 \\
Orderbook OHLCV-Country & \begin{tabular}{@{}rrrr@{}}27.93 & \textbf{0.054} & 84.77 & 226.34\end{tabular} & \begin{tabular}{@{}rrrr@{}}30.51 & 0.040 & 91.67 & 259.80\end{tabular} & \begin{tabular}{@{}rrrr@{}}32.22 & \textbf{0.056} & 93.91 & 263.71\end{tabular} & 2/12 \\
\midrule
Orderbook LMP-Self & \begin{tabular}{@{}rrrr@{}}28.62 & \textbf{0.049} & 85.50 & 250.31\end{tabular} & \begin{tabular}{@{}rrrr@{}}30.48 & 0.067 & 87.37 & 260.43\end{tabular} & \begin{tabular}{@{}rrrr@{}}31.88 & \textbf{0.070} & 89.00 & 263.45\end{tabular} & 2/12 \\
Orderbook LMP-Neighbor & \begin{tabular}{@{}rrrr@{}}32.57 & 0.083 & 94.95 & 249.92\end{tabular} & \begin{tabular}{@{}rrrr@{}}32.95 & 0.086 & 95.31 & 269.34\end{tabular} & \begin{tabular}{@{}rrrr@{}}33.64 & \textbf{0.081} & 96.44 & 266.04\end{tabular} & 1/12 \\
Orderbook LMP-Country & \begin{tabular}{@{}rrrr@{}}28.56 & 0.064 & 83.43 & 243.69\end{tabular} & \begin{tabular}{@{}rrrr@{}}30.16 & 0.053 & 86.49 & 258.76\end{tabular} & \begin{tabular}{@{}rrrr@{}}31.38 & \textbf{0.051} & 88.57 & 262.37\end{tabular} & 1/12 \\
\midrule
Orderbook VWAP-Self & \begin{tabular}{@{}rrrr@{}}27.04 & 0.077 & 82.10 & 218.73\end{tabular} & \begin{tabular}{@{}rrrr@{}}29.00 & 0.072 & 85.72 & 251.60\end{tabular} & \begin{tabular}{@{}rrrr@{}}30.78 & \textbf{0.067} & 88.72 & 256.20\end{tabular} & 1/12 \\
Orderbook VWAP-Neighbor & \begin{tabular}{@{}rrrr@{}}\textbf{25.88} & \textbf{0.056} & \textbf{77.51} & \textbf{198.01}\end{tabular} & \begin{tabular}{@{}rrrr@{}}\textbf{28.43} & 0.054 & \textbf{84.97} & \textbf{245.92}\end{tabular} & \begin{tabular}{@{}rrrr@{}}\textbf{29.68} & \textbf{0.060} & \textbf{86.43} & \textbf{250.32}\end{tabular} & \textbf{11/12} \\
Orderbook VWAP-Country & \begin{tabular}{@{}rrrr@{}}26.75 & \textbf{0.053} & 81.88 & 223.60\end{tabular} & \begin{tabular}{@{}rrrr@{}}28.65 & 0.056 & \textbf{85.15} & 254.87\end{tabular} & \begin{tabular}{@{}rrrr@{}}30.60 & \textbf{0.051} & 88.85 & 260.17\end{tabular} & 3/12 \\
\bottomrule
\end{tabular}%
}
\end{table*}

\begin{table*}[!t]
\caption{Orderbook testing performance across the horizons in Austria.}
\label{tab:orderbook_horizon_at}
\centering
\setlength{\arrayrulewidth}{\lightrulewidth}
\renewcommand{\arraystretch}{1.04}
\makebox[\textwidth][c]{%
\begin{tabular}{lc|c|c|c}
\toprule
Method & \begin{tabular}{@{}rrrr@{}}\multicolumn{4}{c}{\uline{ Horizon $h$ = 15 min}}\\AQL & AQCE & MAE & RMSE\end{tabular} & \begin{tabular}{@{}rrrr@{}}\multicolumn{4}{c}{\uline{ Horizon $h$ = 60 min}}\\AQL & AQCE & MAE & RMSE\end{tabular} & \begin{tabular}{@{}rrrr@{}}\multicolumn{4}{c}{\uline{ Horizon $h$ = 180 min}}\\AQL & AQCE & MAE & RMSE\end{tabular} & Count \\
\midrule
Orderbook OHLCV-Self & \begin{tabular}{@{}rrrr@{}}29.40 & \textbf{0.045} & 77.60 & 347.22\end{tabular} & \begin{tabular}{@{}rrrr@{}}29.97 & 0.035 & 78.71 & 354.65\end{tabular} & \begin{tabular}{@{}rrrr@{}}32.01 & 0.039 & 83.13 & 359.38\end{tabular} & 1/12 \\
Orderbook OHLCV-Neighbor & \begin{tabular}{@{}rrrr@{}}27.50 & \textbf{0.027} & 71.87 & \textbf{333.68}\end{tabular} & \begin{tabular}{@{}rrrr@{}}28.63 & \textbf{0.018} & 74.80 & \textbf{349.43}\end{tabular} & \begin{tabular}{@{}rrrr@{}}30.05 & \textbf{0.030} & 79.07 & 356.83\end{tabular} & 5/12 \\
Orderbook OHLCV-Country & \begin{tabular}{@{}rrrr@{}}30.88 & \textbf{0.037} & 81.91 & 349.61\end{tabular} & \begin{tabular}{@{}rrrr@{}}31.97 & 0.032 & 83.74 & 357.12\end{tabular} & \begin{tabular}{@{}rrrr@{}}33.85 & \textbf{0.028} & 88.59 & 361.91\end{tabular} & 2/12 \\
\midrule
Orderbook LMP-Self & \begin{tabular}{@{}rrrr@{}}31.80 & \textbf{0.096} & 83.30 & 346.63\end{tabular} & \begin{tabular}{@{}rrrr@{}}32.60 & 0.102 & 84.97 & 358.41\end{tabular} & \begin{tabular}{@{}rrrr@{}}33.36 & 0.108 & 86.80 & 362.13\end{tabular} & 1/12 \\
Orderbook LMP-Neighbor & \begin{tabular}{@{}rrrr@{}}27.22 & \textbf{0.058} & 70.91 & \textbf{331.60}\end{tabular} & \begin{tabular}{@{}rrrr@{}}28.59 & 0.053 & 74.19 & \textbf{348.85}\end{tabular} & \begin{tabular}{@{}rrrr@{}}29.93 & 0.052 & 78.67 & 358.83\end{tabular} & 3/12 \\
Orderbook LMP-Country & \begin{tabular}{@{}rrrr@{}}31.19 & \textbf{0.065} & 81.53 & 352.21\end{tabular} & \begin{tabular}{@{}rrrr@{}}31.93 & 0.059 & 82.80 & 357.90\end{tabular} & \begin{tabular}{@{}rrrr@{}}33.28 & 0.075 & 85.22 & 360.60\end{tabular} & 1/12 \\
\midrule
Orderbook VWAP-Self & \begin{tabular}{@{}rrrr@{}}28.81 & \textbf{0.067} & 74.95 & 349.01\end{tabular} & \begin{tabular}{@{}rrrr@{}}29.60 & 0.062 & 77.37 & 354.10\end{tabular} & \begin{tabular}{@{}rrrr@{}}31.72 & 0.058 & 81.78 & 358.43\end{tabular} & 1/12 \\
Orderbook VWAP-Neighbor & \begin{tabular}{@{}rrrr@{}}\textbf{26.85} & \textbf{0.057} & \textbf{69.12} & \textbf{332.64}\end{tabular} & \begin{tabular}{@{}rrrr@{}}\textbf{28.15} & 0.059 & \textbf{72.98} & \textbf{347.55}\end{tabular} & \begin{tabular}{@{}rrrr@{}}\textbf{29.43} & 0.053 & \textbf{76.93} & \textbf{355.05}\end{tabular} & \textbf{10/12} \\
Orderbook VWAP-Country & \begin{tabular}{@{}rrrr@{}}29.06 & \textbf{0.064} & 75.62 & 348.02\end{tabular} & \begin{tabular}{@{}rrrr@{}}29.82 & 0.057 & 77.49 & 354.61\end{tabular} & \begin{tabular}{@{}rrrr@{}}31.92 & 0.058 & 82.36 & 358.79\end{tabular} & 1/12 \\
\bottomrule
\end{tabular}%
}
\end{table*}

\textbf{AQCE.} AQCE measures the absolute difference between the empirical coverage of the prediction interval and its nominal coverage of 0.80. It is defined as
\begin{equation}
\operatorname{AQCE}
=
\left|
\frac{1}{N}
\sum_{i=1}^{N}
\mathbb{I}
\left\{
\widehat{y}_i^{(0.1)}
\leq y_i \leq
\widehat{y}_i^{(0.9)}
\right\}
-0.80
\right|.
\label{eq:aqce}
\end{equation}
A lower AQCE indicates better calibration of the prediction interval.

\textbf{MAE.} MAE evaluates pointwise prediction accuracy. A lower MAE indicates a smaller average absolute prediction error.

\textbf{RMSE.} RMSE also evaluates pointwise prediction accuracy. RMSE assigns greater weight to large prediction errors. A lower RMSE indicates better pointwise prediction performance.

\textbf{DM Test.} The paired DM test determines whether two models have statistically different predictive performance~\cite{diebold2015comparing}. Let $d_i$ be the difference between the losses of two models on sample $i$. The null hypothesis is equal expected loss, $H_0\!:\mathbb{E}[d_i]=0$. A two-sided $p$-value is used to evaluate the null hypothesis. We reject $H_0$ when $p<0.05$, in which case the model with the lower average loss is considered significantly better. Otherwise, the two models are regarded as statistically indistinguishable. One, two, and three stars denote $p<0.05$, $p<0.01$, and $p<0.001$, respectively.

\begin{figure*}[!t]
\centering
\includegraphics[width=\textwidth]{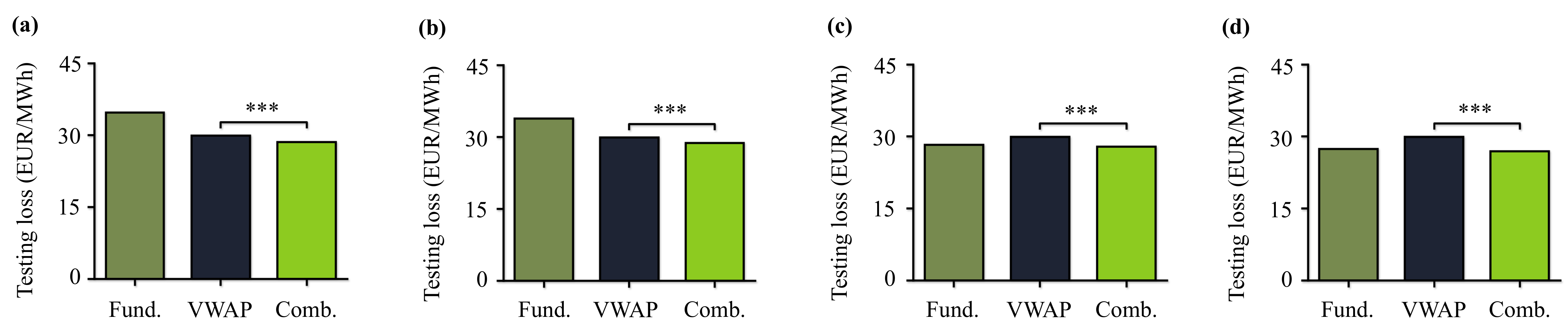}
\caption{Germany testing AQL. Panels \textbf{(a)} to \textbf{(d)} compare Fundamental-1 to Fundamental-4 with the same validation selected best orderbook and their combined model. Each panel contains exactly three bars.}
\label{fig:combined_de}
\end{figure*}

\begin{figure*}[!t]
\centering
\includegraphics[width=\textwidth]{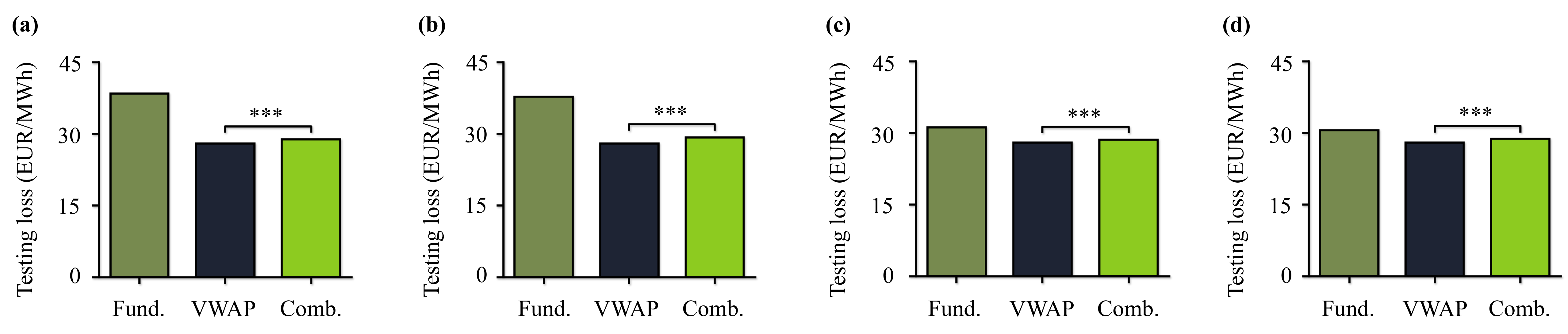}
\caption{Austria testing AQL. Panels \textbf{(a)} to \textbf{(d)} compare Fundamental-1 to Fundamental-4. The three bar and significance conventions follow Fig.~\ref{fig:combined_de}.}
\label{fig:combined_at}
\end{figure*}

We design multiple case studies to identify the optimal orderbook representation in Section~IV-A, investigate the influence of fundamentals and whether orderbook information can replace them in Section~IV-B, and determine the optimal training samples in Section~IV-C. Validation data select hyperparameters. Each model is estimated with three random seeds, and the three folds cover the full 2024 testing year. The rolling origin data splits are reported in Table~\ref{tab:folds}. This design supports evaluation across different seasons \cite{lago2021forecasting}. Results are reported separately for each horizon and country, and each DM test compares two models on the same testing observations. 
For probabilistic evaluation, we use AQL and AQCE. 
For pointwise evaluation, we use MAE and RMSE. Lower values are preferred for every metric.
Moreover, paired DM tests and their $p$ values assess whether loss differences are statistically significant. The stars defined in Section~III-D are reported for the pairwise comparisons in Figs.~\ref{fig:linear_qr_de} to~\ref{fig:combined_at}. In Tables~\ref{tab:orderbook_horizon_de} and~\ref{tab:orderbook_horizon_at}, where nine cases are compared at once, significance is instead conveyed by bold entries.

\begin{figure}[!ht]
\centering
\includegraphics[width=\columnwidth]{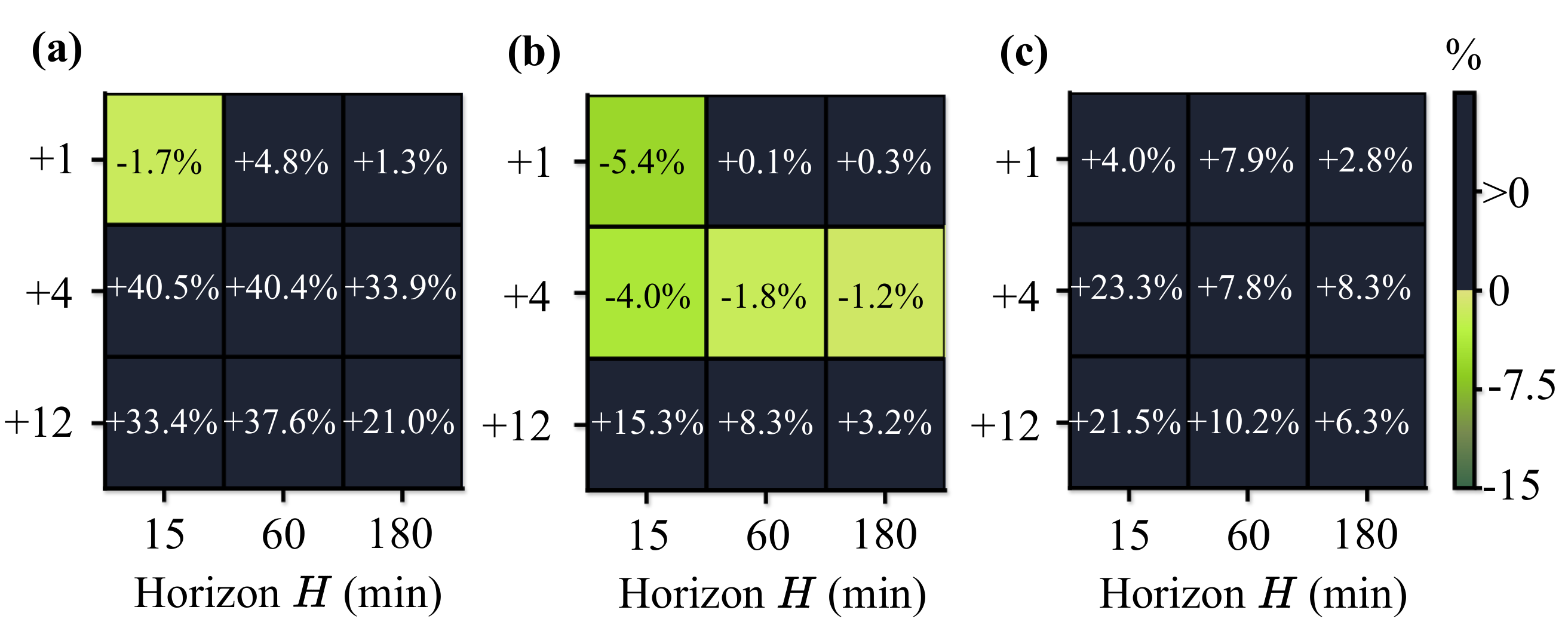}
\caption{Germany percentage change in AQL for the $+1$, $+4$, and $+12$ later product contexts relative to Self. Results are computed from independently trained 15, 60, and 180 minute models and averaged across three seeds and three folds. Panels \textbf{(a)} to \textbf{(c)} show OHLCV, VWAP, and LMP. Nonpositive changes use a dark to light green scale, while positive changes are navy.}
\label{fig:orderbook_context_de}
\end{figure}

\begin{figure}[!ht]
\centering
\includegraphics[width=\columnwidth]{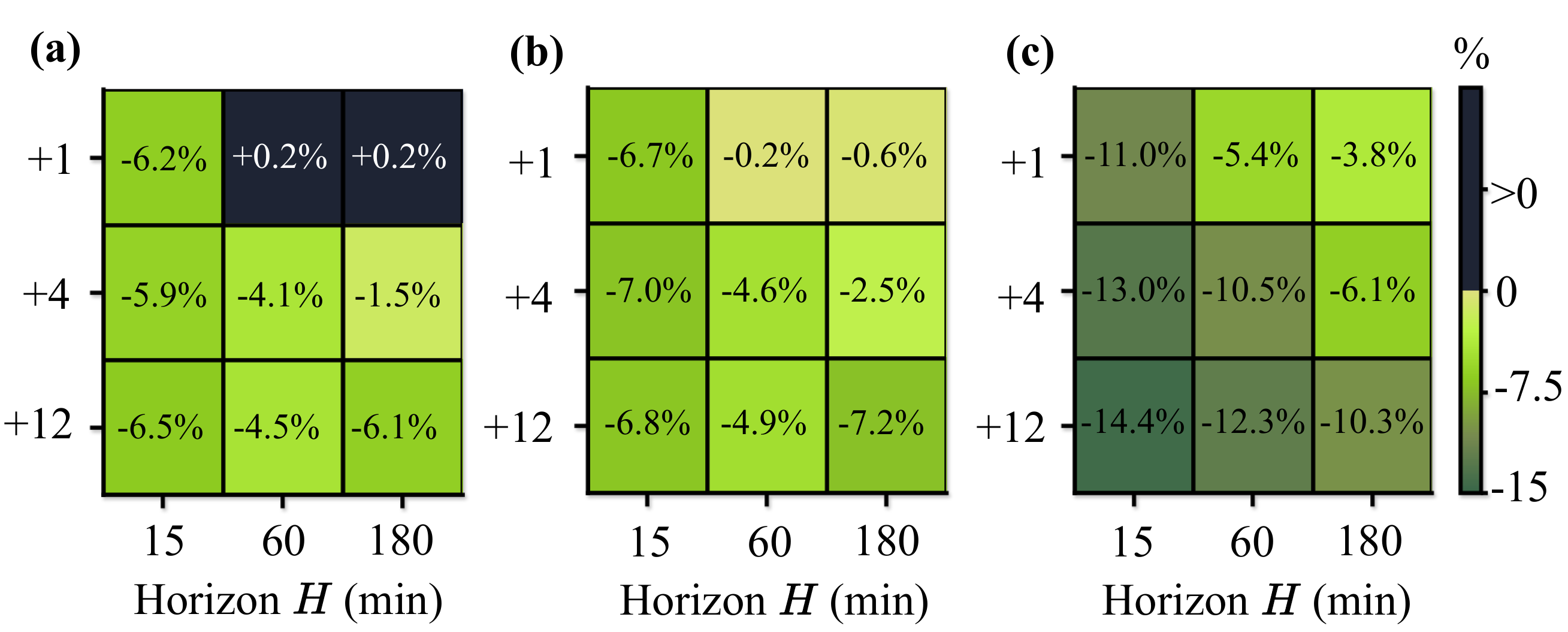}
\caption{Austria percentage change in AQL for the three later product contexts relative to Self. Panel order, horizon categories, color semantics, and aggregation follow Fig.~\ref{fig:orderbook_context_de}.}
\label{fig:orderbook_context_at}
\end{figure}

\subsection{Representation of Orderbook}
 \textbf{Setting.} We compare OHLCV, LMP, and VWAP. For every representation, we test the self-product delivered at the modeling target (suffix -Self), later neighboring products anchored to that same target (suffix -Neighbor), and the corresponding self-product from the neighboring country (suffix -Country), as defined in \eqref{eq:featuresets}.

\begin{figure*}[!t]
\centering
\includegraphics[width=\textwidth]{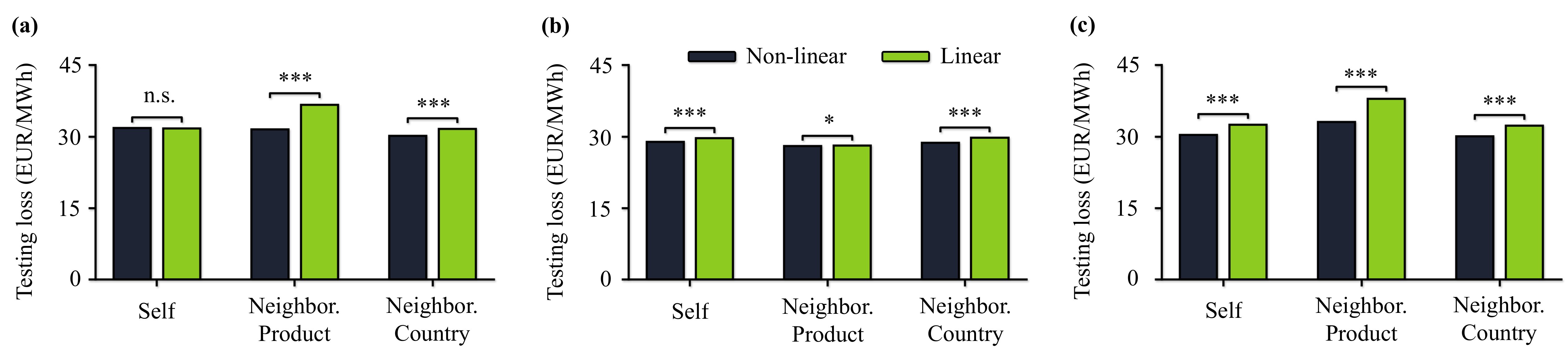}
\caption{Germany ANN and LQR testing AQL averaged across the independently trained 15, 60, and 180 minute targets and three folds. Panels \textbf{(a)} to \textbf{(c)} show OHLCV, VWAP, and LMP.}
\label{fig:linear_qr_de}
\end{figure*}

\begin{figure*}[!t]
\centering
\includegraphics[width=\textwidth]{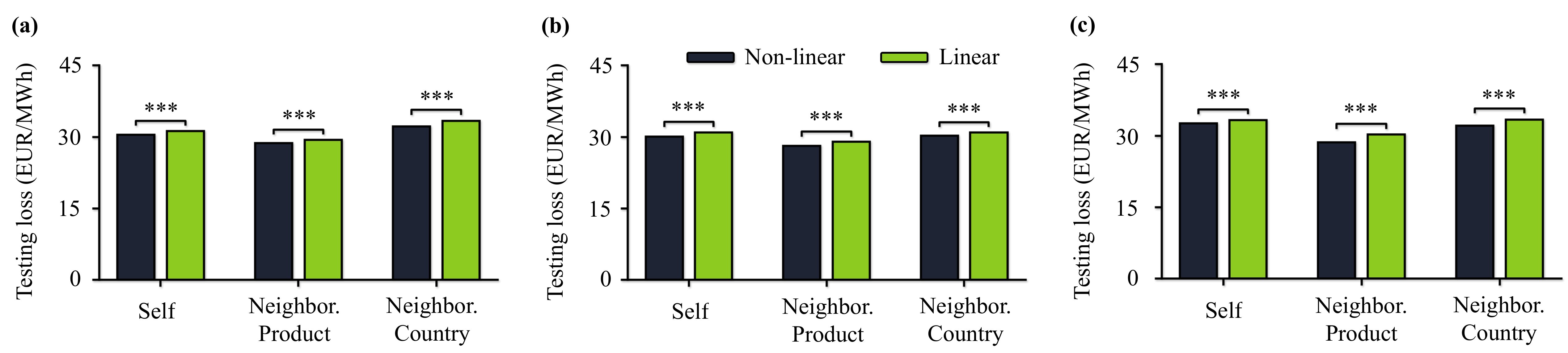}
\caption{Austria ANN and LQR testing AQL for the nine orderbook cases. Panel order, aggregation, and significance conventions follow Fig.~\ref{fig:linear_qr_de}.}
\label{fig:linear_qr_at}
\end{figure*}

 \textbf{Results.} Tables~\ref{tab:orderbook_horizon_de} and~\ref{tab:orderbook_horizon_at} compare the nine representation cases. Bold entries belong to the statistically strongest group. Multiple bold entries are statistically indistinguishable under paired Diebold-Mariano tests with $p>0.05$. The Count column reports how many of the 12 metric and horizon cells belong to that group. VWAP with neighboring products (VWAP-Neighbor) is consistently the strongest case in Germany and Austria, with counts of 11 and 10. 
 Here, strongest is defined through the Count column. 
 
 Four observations follow from Tables~\ref{tab:orderbook_horizon_de} and~\ref{tab:orderbook_horizon_at}. First, VWAP-Neighbor has the lowest AQL, MAE, and RMSE at every horizon in both countries. Second, the loss of every representation increases with the horizon, as the orderbook at the origin is less informative about a more distant delivery. Third, the neighboring country adds modest value in Germany, where OHLCV-Country and VWAP-Country improve on their self-product counterparts, but no value in Austria. Fourth, AQCE is small for all cases, so the representations differ in accuracy rather than in calibration. An intuition for the dominance of VWAP-Neighbor is that VWAP summarizes the volume-weighted consensus price of each interval in a single value that is robust to small orders, whereas LMP uses only two observations and OHLCV multiplies the input dimension with extreme prices that are often set by small orders. Neighboring products add information as the continuous intraday market allows traders to trade several products in parallel, and traders split large orders both over the historical window and across neighboring products to reduce price impact. The orderbooks of neighboring products therefore carry part of the same position that is being closed for the target product.
 Fig.~\ref{fig:orderbook_context_de} and Fig.~\ref{fig:orderbook_context_at} show the AQL change from adding each later product context. For Austrian VWAP and LMP, every added context reduces testing loss. For Germany, adding later products to OHLCV or LMP generally increases testing loss, while the VWAP response depends on the amount of context and the horizon.

\textcolor{black}{Moreover, compared with Austria, Germany exhibits smaller and less consistent gains from neighboring product information, suggesting that predictive orderbook information is more localized around the target product. One possible explanation is that Austria is a smaller market, where traders may place bids for multiple delivery products at the same time, which could cause predictive information to be distributed more broadly across neighboring products.} An alternative explanation is liquidity: the orderbook of a single Austrian product contains roughly ten times fewer trades than a German one, so its features are noisier and averaging over neighboring products acts as denoising. 
Fig.~\ref{fig:linear_qr_de} and Fig.~\ref{fig:linear_qr_at} compare ANN with LQR for the same nine cases. ANN has significantly lower loss in all nine Austrian cases and eight of the nine German cases. German OHLCV with Self is statistically indistinguishable. The results indicate that interactions across neighboring products and countries are generally not captured as effectively by a linear specification. Hence, a nonlinear model is needed. In summary, Figs.~\ref{fig:orderbook_context_de} to~\ref{fig:linear_qr_at} yield three takeaways. In Austria, every added neighboring product reduces the loss for VWAP and LMP, and the reduction grows with the number of products, up to 14\% for LMP with 12 products. In Germany, neighboring products only help VWAP, and adding 4 or 12 products to OHLCV increases the loss by up to 40\%, so the amount of context has to be validated per representation. Finally, the ANN advantage over LQR is largest for the neighboring product cases, so the value of the added context is only realized by a nonlinear model.

\begin{figure*}[!t]
\centering
\includegraphics[width=\textwidth]{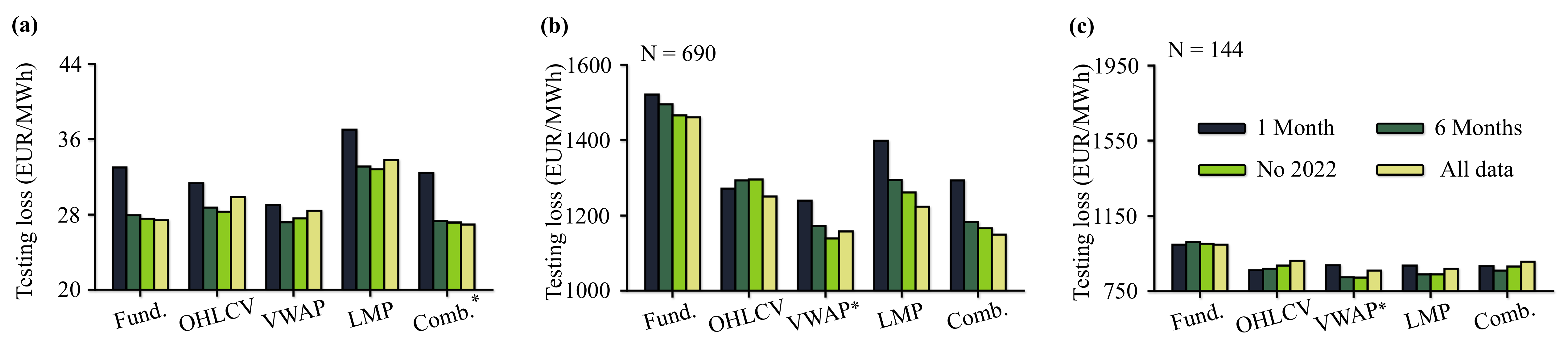}
\caption{Germany testing AQL for the 1-month, 6-month, No-2022, and all-data training samples. Panel \textbf{(a)} uses all testing observations. Panels \textbf{(b)} and \textbf{(c)} use observations with realized imbalance prices above 1000 EUR/MWh and below $-1000$ EUR/MWh. $N$ is the number of qualifying sample and horizon pairs.}
\label{fig:training_history_de}
\end{figure*}

\begin{figure*}[!t]
\centering
\includegraphics[width=\textwidth]{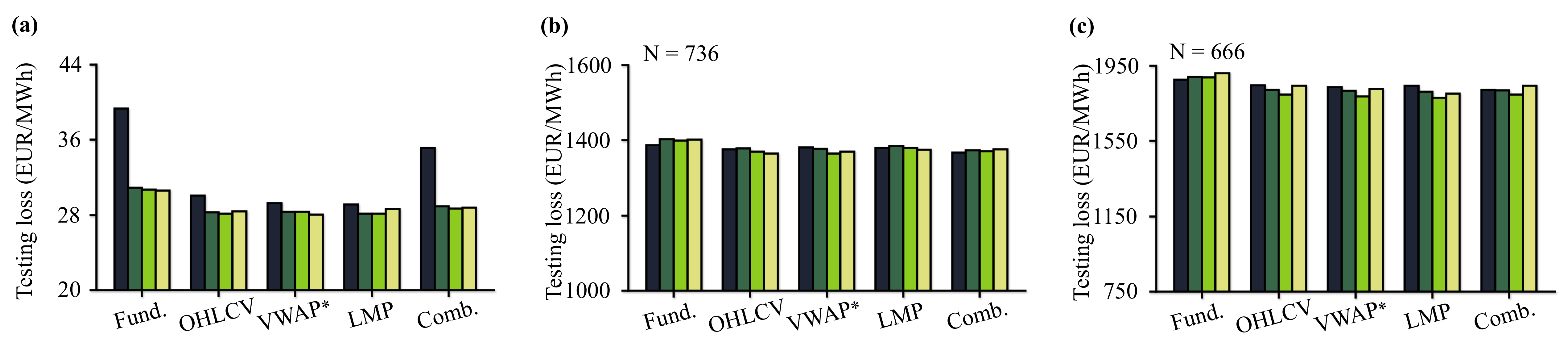}
\caption{Austria overall and extreme event comparisons across training histories. Panel order, aggregation, colors, significance marks, and the definition of $N$ follow Fig.~\ref{fig:training_history_de}.}
\label{fig:training_history_at}
\end{figure*}

\subsection{Fundamentals versus Orderbook}
 \textbf{Setting.} We fix the strongest VWAP specification selected on validation data in each country and fold. Fundamental-1 to Fundamental-4 are considered separately using the cumulative feature sets in Table~\ref{tab:fundamental_sets}. For each set, we estimate a fundamental model, compare it with the selected orderbook model, and combine the two inputs in a third model. If combination does not improve on the orderbook model, it indicates that the orderbook already carries sufficient information from the examined fundamentals for this modeling task.

 \textbf{Results.} Fig.~\ref{fig:combined_de} and Fig.~\ref{fig:combined_at} isolate the incremental value of Fundamental-1 to Fundamental-4. Each panel compares the fundamental set, the validation-selected orderbook, and their combined model. In Germany, combining VWAP with fundamentals consistently reduces testing loss relative to VWAP alone. This indicates that the German orderbook does not fully reflect the useful information in the fundamental inputs. In Austria, adding fundamentals to VWAP consistently increases testing loss. This indicates that the Austrian orderbook is sufficient for the examined task and that the added fundamental features provide redundant or noisy information. These are predictive findings on the tested feature sets and models.
\textcolor{black}{Furthermore, the results suggest that renewable generation features play an important role in Germany, while cross-border nominations provide further value.} The country difference is consistent with the following explanation, which the loss comparisons do not identify directly. In the small Austrian market, imbalance prices are driven more by positions left open after intraday trading, which the orderbook reflects directly, whereas in Germany the large share of wind and solar generation creates physical uncertainties after gate closure that the orderbook cannot fully price in.

\subsection{Optimal Training Samples}
 \textbf{Setting.} To isolate training history effects, validation and testing periods remain unchanged. For the selected fundamental, OHLCV, VWAP, LMP, and validation-selected combined configurations, we compare the latest 1 month, the latest 6 months, all available observations after 2022, which is denoted by No 2022, and the original expanding sample with all available observations.

 \textbf{Results.} Fig.~\ref{fig:training_history_de} and Fig.~\ref{fig:training_history_at} compare the all data benchmark with each restricted history model for Germany and Austria. Panel \textbf{(a)} averages AQL across all testing observations, while panels \textbf{(b)} and \textbf{(c)} condition on positive and negative extreme prices. The comparison keeps the validation and testing windows fixed, so differences reflect only the training sample. The German testing set contains 690 positive and 144 negative extreme sample-horizon pairs, while the Austrian set contains 736 and 666. Although Germany benefits from combining orderbook and fundamentals overall, VWAP alone gives the lowest loss for its positive and negative extreme observations. For overall performance, the strongest German and Austrian cases use all available data. For the extreme observations, \textcolor{black}{excluding 2022 data} gives the lowest VWAP loss in both countries. This ablation does not establish that energy crisis data should always be excluded for extreme price modeling. The appropriate sample depends on whether future conditions are expected to resemble the 2022 crisis.

\section{Limitations and Future Work}
Several limitations remain, and they bound the scope of our experimental settings. First, the analysis is limited to Germany and Austria as the orderbook data are commercial. Future work should examine other markets, in particular markets with different imbalance pricing rules, generation mixes, and liquidity, to assess how broadly the country-specific conclusions apply. Second, the study considers three orderbook representations, and future work can evaluate richer representations while retaining the same market interaction framework, for example order-level depth features or end-to-end orderbook encoders \cite{yu2026orderfusionencodingorderbookendtoend}, and additional fundamental features beyond Table~\ref{tab:fundamental_sets} might further improve the combined models. Third, our evidence is predictive and does not identify the causal mechanism of the interaction, which calls for structural or causal analysis. In particular, the explanations offered for the country differences, namely open trading positions in Austria and physical uncertainties in Germany, and for the neighboring product gains are consistent with the loss comparisons but are not identified by them. Fourth, the economic value of orderbook-based imbalance price models in trading and storage bidding remains to be quantified. Lastly, it is worth continuously monitoring whether the optimal training sample size varies, especially when new electricity market regulations or policies are introduced.

\section{Conclusion}

Intraday and balancing markets are usually studied separately, although positions left open in the former shape prices in the latter. 
This paper provides a systematic analysis of cross-market interaction from the orderbook perspective by examining how continuous intraday orderbook information should be represented for probabilistic imbalance price modeling in Germany and Austria.
Out of OHLCV, VWAP, and LMP, we found that VWAP with later product context was the strongest orderbook representation in both countries, and the nonlinear ANN significantly outperformed LQR in 17 of the 18 cases. Thus, a nonlinear model is needed. Furthermore, the orderbook alone is sufficient in Austria but not in Germany, where adding fundamentals consistently reduces the loss. \textcolor{black}{The country comparison suggests two patterns in imbalance price formation. The dominance of VWAP in Austria is consistent with a greater role for open trading positions, whereas the value added by fundamentals in Germany suggests a greater role for physical uncertainties.} Moreover, \textcolor{black}{using all available training data gave the best performance across the full testing set, whereas excluding 2022 reduced VWAP loss specifically for extreme price observations.} Although limited to two countries, these case studies show that a single real-time market signal carries much of the information behind imbalance price formation, and that the extent to which the intraday market prices in fundamentals is country-dependent.

\section*{Appendix}
\subsection{Data Sources}
\textcolor{black}{The continuous intraday orderbook data used in this study are commercial EPEX SPOT market data and can be purchased through the EEX Group Webshop at \url{https://webshop.eex-group.com/} under the EPEX SPOT public market data products. The orderbook data are also available in real time through the corresponding API services. The fundamental features are obtained from the ENTSO-E Transparency Platform at \url{https://transparency.entsoe.eu/}. Some fundamental variables are subject to irregular publication delays.
These variables are therefore not reliably available at every modeling origin, whereas the orderbook is observed in real time. }

\subsection{Hyperparameters}
Table~\ref{tab:hyper} lists the search ranges of the ANN hyperparameters.

\begin{table}[h]
\caption{ANN hyperparameter search ranges.}
\label{tab:hyper}
\centering
\scriptsize
\setlength{\tabcolsep}{2.4pt}
\begin{tabular*}{\columnwidth}{@{\extracolsep{\fill}}ll}
\toprule
Hyperparameter & Search range \\
\midrule
Number of neurons per layer & 32, 64, 128, 256 \\
Number of hidden layers & 2, 3, 4 \\
Activation function & ReLU, Swish \\
Learning rate & $4\times10^{-4}$, $1\times10^{-3}$, $4\times10^{-3}$ \\
\bottomrule
\end{tabular*}
\end{table}

\bibliographystyle{IEEEtran}
\bibliography{references}

\end{document}